# Exciton diffusion in a hBN-encapsulated monolayer MoSe$_2$

Takato Hotta[1], Shohei Higuchi[1], Akihiro Ueda[1], Keisuke Shinokita[2], Yuhei Miyauchi, Kazunari Matsuda[2], Keiji Ueno[3], Takashi Taniguchi[4], Kenji Watanabe[5], and Ryo Kitaura[1,*]

[1]*Department of Chemistry, Nagoya University, Nagoya 464-8602, Japan*

[2]*Institute of Advanced Energy, Kyoto University, Uji, Kyoto 611-0011, Japan*

[3]*Department of Chemistry, Saitama University, Saitama 338-8570, Japan*

[4]*International Center for Materials Nanoarchitectonics, National Institute for Materials Science, 1-1 Namiki, Tsukuba 305-0044, Japan*

[5]*Research Center for Functional Materials, National Institute for Materials Science, 1-1 Namiki, Tsukuba 305-0044, Japan*

*Corresponding authors E-mail: r.kitaura@nagoya-u.jp

Excitons, quasi particles composed of an electron and a hole, play an important role in optical responses in low-dimensional nanostructures. In this work, we have investigated exciton diffusion in a monolayer MoSe$_2$ encapsulated between flakes of hexagonal boron nitrides (hBN/MoSe$_2$/hBN). Through PL imaging and numerical solving the 2D diffusion equation, we revealed that temperature dependence of exciton mobility, $\mu_{\text{ex}}$, in the hBN/MoSe$_2$/hBN shows non-saturating increase at low temperature, which is qualitatively different from those of quantum wells composed of compound semiconductors. Ultraflat structure of monolayer MoSe$_2$ in the hBN/MoSe$_2$/hBN probably leads to the suppression of charged-impurity scattering and surface-roughness scattering.

Excitons are quasi particles composed of an electron and a hole. In response to optical excitations of semiconductors, electrons and holes are generated, leading to formation of hydrogen-like bound states, called excitons, through columbic interactions. In a typical semiconductor such as silicon, energy scale of binding energy between an electron and a hole is small (Si: 14.7 meV, room temperature: 25.8 meV), and in this case, excitonic effect is not dominant in optical responses at room temperature[1]. At cryogenic temperature, however, excitonic effect gives sharp resonances at energies lower than that of the band edge, changing optical spectrum drastically.

In low-dimensional materials, excitonic effect can dominate optical responses even at room temperature[2, 3]. Due to the strong coulomb interaction arising from the reduced dimension, exciton binding energy can reach several hundreds of meV in low-dimensional materials. For example, spectroscopic characterization and theoretical analyses have revealed that two-dimensional (2D) materials, including monolayer $MoS_2$, $WS_2$, and $MoSe_2$, possess exciton binding energy of several hundreds of meV, which is much larger than that of the thermal energy at room temperature[4, 5]. Due to this large binding energy, absorption and photoluminescence (PL) spectra of 2D materials are totally dominated by the excitonic effect, showing strong exciton resonances even at room temperature[6, 7, 23-25]. The excitonic effect is, thus, essential to understand optical properties of 2D materials.

In optical response of 2D materials, exciton diffusion plays an important role. Because exciton binding energy is much larger than the thermal energy at room temperature, optical excitation inevitably leads to formation of excitons in 2D materials. Due to finite kinetic energy of generated excitons, excitons diffuse along a 2D plane before radiative or non-radiative recombination. The three processes, generation, diffusion, and recombination of excitons, always occur in response to optical excitations of 2D materials, and hence, it is important to understand exciton diffusion to understand optical properties of 2D materials. In addition, understanding of exciton diffusion gives basis to develop novel optoelectronics, excitonics, such as excitonic transistors, where control of movement of excitons plays an essential role[8, 26-28]. Although a lot of works on excitons generation and recombination have been done in the past 10 years, research focused on exciton diffusion is still sparse.[9-12]

In this work, we have experimentally investigated the exciton diffusion in a 2D semiconductor, monolayer $MoSe_2$. To suppress unwanted environmental effects arising from substrates and adsorbates, we encapsulated a monolayer $MoSe_2$ by flakes of hexagonal boron nitrides (hBN). The hBN-encapsulation yields ultraflat monolayer $MoSe_2$ that is apart from scattering sources in substrates[13, 14]. PL spectra of hBN/$MoSe_2$/hBN give intense peak arising from radiative recombination of excitons, whereas peak intensity from trions is weak. PL imaging has clearly shown that excitons diffuse before recombination; bright region in PL images are broader than corresponding laser spots used to excite samples. Detailed analyses based on solving the two-dimensional diffusion equation yielded that mobility of excitons increases as temperature decreases, revealing that scattering is strongly suppressed in the hBN-encapsulated samples because of the ultraflat structure and charge neutrality of excitons

We fabricated a monolayer MoSe$_2$ sandwiched between hBN flakes, hBN/MoSe$_2$/hBN, by the dry-transfer method; a monolayer MoSe$_2$ exfoliated on SiO$_2$/Si was picked up with a flake of hBN and transferred onto another flake of hBN[15, 16]. Figure 1a shows an optical microscope image of the fabricated hBN/MoSe$_2$/hBN. Green and blue contrasts correspond to hBN/MoSe$_2$/hBN and a SiO$_2$/Si substrate, respectively. The corresponding PL image measured at 300 K gives uniform intensity all over the sample, indicating that there is no bubble or impurities encapsulated between hBN and MoSe$_2$. Although there are impurities that probably arise from impurities attach on the surface, there is no bubble-like structure observed in an AFM image of the hBN/MoSe$_2$/hBN (Figure 1c and 1d), which is consistent to the uniform intensity observed in the PL image shown in Figure 1b. In this work, we used a relatively thick hBN flake, whose thickness is ca. 20 nm, to minimize the adverse substrate effect, which arises from charged impurities and surface roughness of SiO$_2$/Si.

PL spectrum measured at room temperature gives broad single peak, which originates from radiative recombination of bright excitons, at room temperature. Figure 2a shows temperature dependence of PL spectra of hBN/MoSe$_2$/hBN. As you can see, the PL peak shifts toward the blue side as temperature decreases, which shift can be interpreted by the temperature dependent change in bandgap of MoSe$_2$; the temperature dependent bandgap change have been well reproduced by the Varshini's equation[17], giving parameters of $E_g$ = 1.64 eV, $\alpha$ = 5.07 × 10$^4$ eV K$^{-1}$ and $\beta$ = 2.99 × 10$^2$ K (see Supplementary information)[29 - 31]. The peak from radiative recombination of charged excitons (trions), which locates at the energy 27 meV lower than that from excitons[34], is weak in this sample. This weak peak from trions means that unintentional carrier doping from the substrate and impurities is very small. Figure 2b shows temperature dependence in homogeneous linewidth (Lorentzian width), which was determined through peak fittings with Voigt function; the Lorentzian linewidth was extracted whereas the Gaussian linewidth was fixed at the value determined at 10 K (1.2 meV). As you can see, the values of linewidth become narrower as temperature decreases. We have fitted the temperature dependence of homogeneous linewidth with the following equation, where the first, second, and third term corresponds to residual linewidth, acoustic phonon, and optical phonon scattering, respectively.

$$\gamma = \gamma_0 + c_1 T + \frac{c_2}{\exp\left(\frac{\Omega}{kT}\right) - 1}$$

$\gamma_0$, $c_1$, $c_2$, $\Omega$, $k$, and $T$ in this equation represent residual linewidth, the constant for acoustic phonon, the constant for optical phonon, optical phonon energy, Boltzmann constant and temperature, respectively. The fitting yields the parameters of $\gamma_0$ = 2.8 ± 0.5 meV, $c_1$ = 20 ± 3 μeV K$^{-1}$, $c_2$ = 73 ± 1 meV and $\Omega$ = 32 ± 3 meV. As clearly demonstrated, the contribution from optical phonon is dominant at temperature higher than about 100 K, and at low temperature, linewidth decreases linearly against temperature, which means acoustic phonon scattering is dominant at low temperature. The obtained value of $\Omega$ is consistent to values in previous reports[32), 33)].

We have investigated exciton diffusion based on the PL imaging technique. Figure 3a and b show an image of the laser spot used to excite the sample and the corresponding PL image. We use wavelength of 633 nm to excite the sample and a long-pass filter (650 nm) to filter out reflected light to form PL images. We have investigated excitation power dependence on PL intensity before the PL imaging measurements, and excitation power, which is well below the onset of exciton annihilation, was used for all measurements to avoid formation of hot excitons[21, 22)]. As seen in the line profile shown in Figure 3c, the PL image is broader than the image of the laser spot. This broadening arises from diffraction limit of light and exciton diffusion; excitons diffuse along the 2D plane before radiative recombination, leading to broadening of PL images. The diffraction limit of light can be modeled with the Gaussian-type point spread function, whose Gaussian sigma, $\sigma_{diff}$, is 0.21 $\lambda/NA$[18)]; we use the center wavelength of exciton emission in PL, $\lambda$ = 756 ~ 792 nm and the value of numerical aperture ($NA$) of 0.7 to evaluate the contribution from the diffraction limit. To evaluate contribution of exciton diffusion quantitatively, we have numerically solved 2D diffusion equation with the exciton decay term.

$$\frac{\partial}{\partial t}N(x,y,t) = D\frac{\partial^2}{\partial x^2}N(x,y,t) + D\frac{\partial^2}{\partial y^2}N(x,y,t) - \frac{1}{\tau}N(x,y,t)$$

$x$, $y$, and $t$ represent $x$ and $y$ coordinates in the 2D plane and time, respectively. $N$, $D$, and $\tau$ correspond to number of excitons, diffusion constant, and lifetime of excitons. We assume that the diffusion of exciton is isotropic and laser spot can be modeled as Gaussian function. To solve the 2D diffusion equation, we need to know exciton lifetime, $\tau$. We experimentally determined $\tau$ through measurements of time-resolved PL intensity with the time-correlated single photon counting (TCSPC) method. In most of the cases, time dependences of PL intensity are fitted with the two-component exponential decay model, where a long-lifetime component and a short-lifetime component simultaneously exist. We, therefore, have added up solutions of two independent diffusion equations with different lifetimes to take both the long-lifetime and the short-lifetime components into account. After solving the diffusion equations, we have summed obtained $N(x, y)$ at different times (from zero to 3$\tau$) to calculate PL intensity distributions; we have convoluted the point-spread function to compare calculated PL intensity to observed one. Figure 3a and 3b shows the line profile of PL image measured at 20 K and the calculated PL profile. As shown in Figure 3c, the calculated PL profile reproduces the observed profile well, yielding the value of $D$ of 21 cm$^2$/s. The diffusion constant can be converted to mobility of excitons with the Einstein relation, $\mu_{ex} = eD/k_BT$, and the room temperature mobility was determined to be $1.2 \times 10^4$ cm$^2$/Vs.

To have insight on exciton diffusion in hBN/MoSe$_2$/hBN more, we have measured temperature dependence in exciton diffusion from 60 to 10 K. Using the measured $\tau$ (Fig. S4), we have performed the numerical solving of the diffusion equation at each temperature, yielding temperature dependence of $\mu_{ex}$ (Figure 4). As seen in the figure, $\mu_{ex}$ increases as temperature decreases, and $\mu_{ex}$ exceeds $10^4$ cm$^2$/Vs at low temperatures. Let us compare the obtained results with those of GaAs-based quantum wells (QWs)[19, 20]. In the case of QWs, it has been reported that $\mu_{ex}$ increases as temperature decreases from room temperature to around 100 K, but as temperature decrease more, $\mu_{ex}$ shows saturation and decrease as temperature goes down to cryogenic temperature. At high temperature region, $\mu_{ex}$ increases as temperature decreases due to suppression of phonon scattering, and the same tendency is seen in hBN/MoSe$_2$/hBN. At cryogenic temperature, however, the difference between QWs and hBN/MoSe$_2$/hBN become obvious. The saturation and decrease in $\mu_{ex}$ in QWs arise from the interface roughness scattering, which dominate scattering process at low temperature; the interface roughness scattering is prominent in the case of thin QWs. In contrast, in spite of the ultra-thin structure of monolayer MoSe$_2$, whose thickness is about 0.7 nm, $\mu_{ex}$ in hBN/MoSe$_2$/hBN shows non-saturating increase throughout temperature investigated, 10 ~ 60 K. This non-saturating increase should originate from atomically flat structure of MoSe$_2$ in hBN/MoSe$_2$/hBN, where the atomically flat hBN flakes ensures the flatness of MoSe$_2$. The atomically flat structure in hBN/MoSe$_2$/hBN strongly suppresses the interfacial roughness scattering, leading to the non-saturating $\mu_{ex}$. QWs and 2D materials are similar systems in terms of 2D electronic systems, but the ultrathin and ultraflat structure of 2D materials gives distinctly different nature to excitons in 2D materials.

Because excitons are neutral objects, scattering from charged impurity should be suppressed in the case of excitons. Previous studies on carrier transport of hBN-encapsulated TMDs have revealed that carrier mobility is enhanced in the case of multilayer structure. For example, X. Cui et al. have reported that a 6-layer MoS$_2$ shows carrier mobility of $3 \times 10^4$ cm$^2$/Vs whereas a monolayer MoS$_2$ shows ~ $10^3$ cm$^2$/Vs at low temperature[13]. They have concluded that the significant difference in mobility arises from interfacial impurity scattering; the thicker MoS$_2$ is the larger distance between interfacial impurity and carriers is. This clearly demonstrates that charged impurity scattering is one of the dominant limiting factors of carrier mobility in TMDs. In contrast, in the case of excitons, charged impurity scattering should be suppressed due to the charge neutrality of excitons, and this might play an important role in the observed large $\mu_{ex}$. Note that absolute value of $\mu_{ex}$ depends strongly on exciton lifetime and determination of precise values of $\mu_{ex}$ needs further measurement with a sample whose lifetime is long enough to minimize experimental error. However, lower bound estimation of $\mu_{ex}$ with upper bound value of exciton lifetimes still ranges the order of ~ $10^3$ cm$^2$/Vs, which should result from the suppression of interfacial impurity scattering.

Suppression of charged impurity scattering can be seen in temperature dependence of linewidth of the excitonic PL peak. As discussed above, at low temperature (< 100 K), linewidth of the excitonic peak decreases linearly against temperature, which means that acoustic phonon scattering is the dominant factor in this temperature regime. This is consistent to the observed temperature dependence in $\mu_{ex}$. Fitting of the observed temperature dependence of $\mu_{ex}$ in Figure 4 gives the relation of $\mu_{ex}(T) \propto T^{-1.2}$. It is well known that carrier mobility limited by acoustic phonon scattering is proportional to $T^{-1}$ in 2D electronic system, which relation is close to the observed relation, $\mu_{ex}(T) \propto T^{-1.2}$. Similar $\mu$-$T$ relation is also seen in the other hBN/MoSe$_2$/hBN sample (Fig. S5), and this strongly suggests that impurity scattering is suppressed in exciton diffusion in a high-quality hBN/MoSe$_2$/hBN sample investigated in this study.

In conclusion, we have investigated exciton diffusion in a hBN/MoSe$_2$/hBN through PL imaging and numerical solving the 2D diffusion equation. Images of laser spots used to excite the sample are smaller than those of corresponding PL images, which means that generated excitons diffuse along the 2D plane before radiative and non-radiative recombination. Detailed analyses based on numerically solving the 2D diffusion equation have yielded exciton mobility, $\mu_{ex}$, at temperature from 10 ~ 60 K. The observed temperature dependence of $\mu_{ex}$ shows non-saturating increase at low temperature, which is significantly different from those of QWs. Ultraflat structure of monolayer MoSe$_2$ in hBN/MoSe$_2$/hBN probably leads to the suppression of charged-impurity scattering and surface-roughness scattering, and the observed temperature dependence of linewidths is consistent to the suppression of scatterings. Our work shows the exciton diffusion characteristic to ultraflat 2D semiconductors, which gives basis to understand basic optical responses in 2D semiconductors.

This work was supported by JSPS KAKENHI Grant numbers JP16H06331, JP16H03825, JP16H00963, JP15K13283, JP25107002, and JST CREST Grant Number JPMJCR16F3. K.W. and T.T. acknowledge support from the Elemental Strategy Initiative conducted by the MEXT, Japan, Grant Number JPMXP0112101001, JSPS KAKENHI Grant Numbers JP20H00354 and the CREST(JPMJCR15F3), JST.

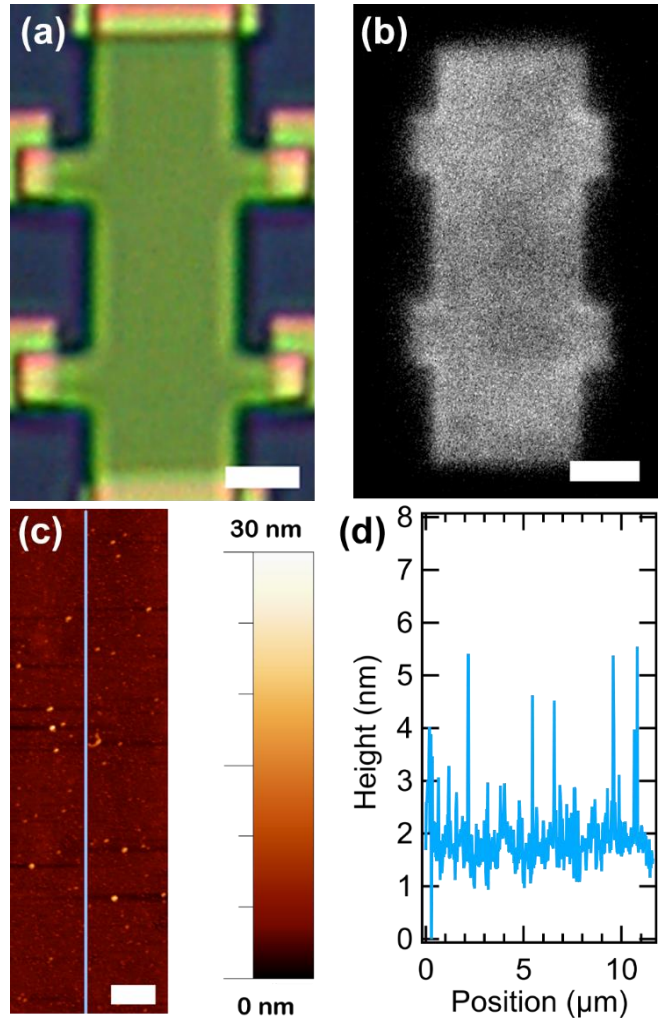

Figure 1(a) An optical microscope image of a hBN-encapsulated monolayer MoSe$_2$ fabricated. Details of the fabrication procedure are shown in the supplementary material. Scale bar, 2 μm. (b) A corresponding PL image. Scale bar, 2 μm. (c) and (d) An AFM image of the hBN-encapsulated monolayer MoSe$_2$ and a line profile along the line in the AFM image. Scale bar, 1 μm.

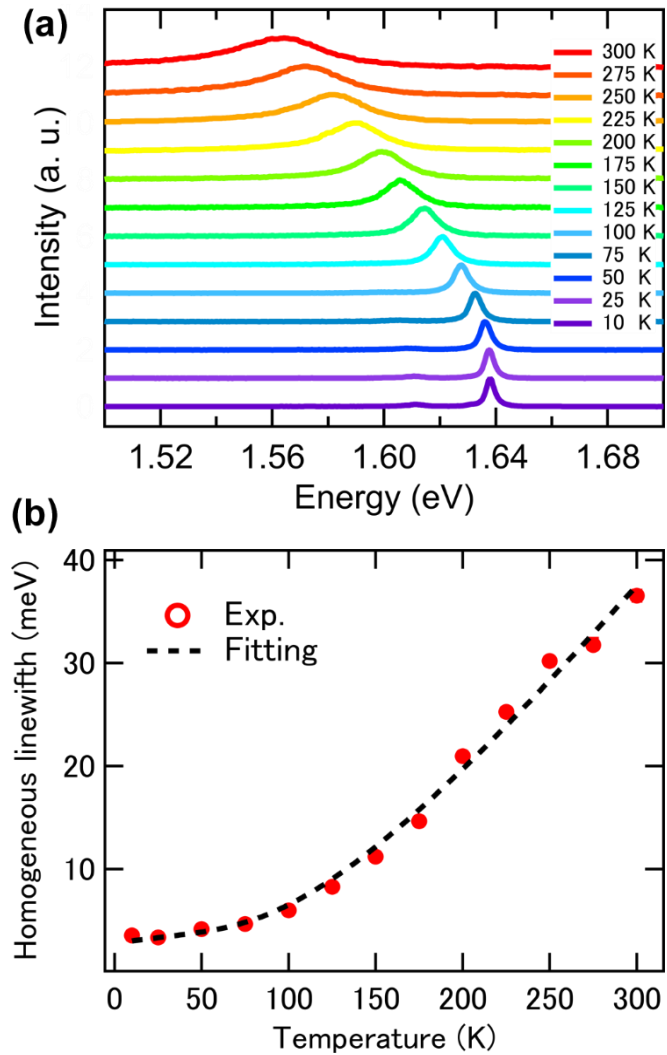

Figure 2 Temperature dependence of PL spectra of a hBN-encapsulated monolayer MoSe$_2$. The excitation wavelength of 550 nm with excitation power of 17 μW was used for the measurements. (b) Temperature dependence of peak positions arising from radiative recombinations of excitons. The dotted line corresponds to the fitting with the equation shown in the main text.

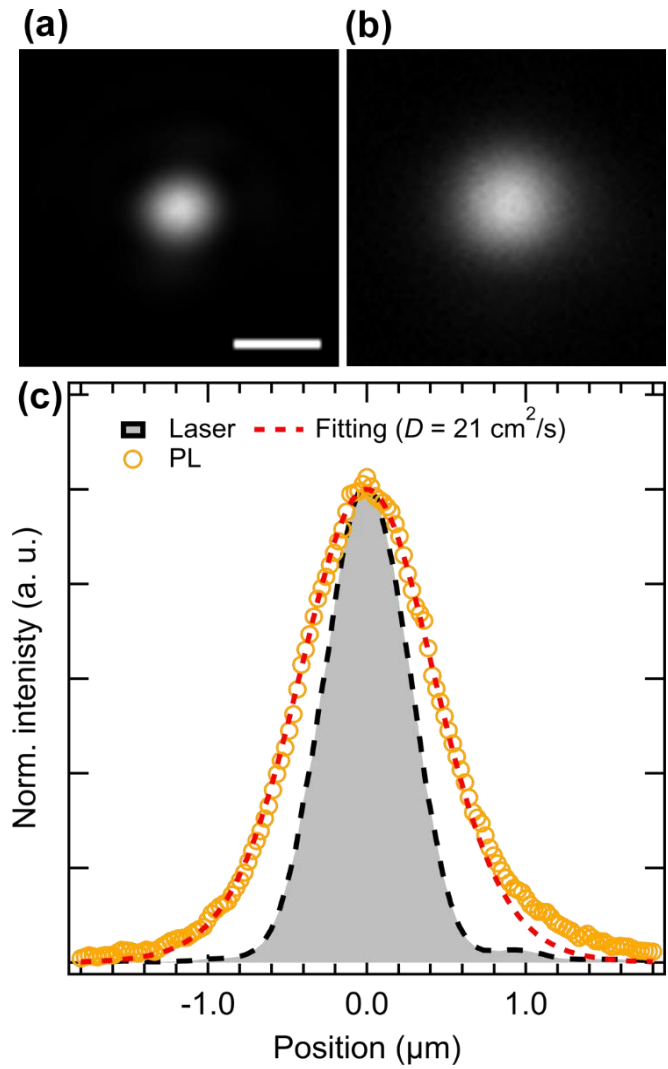

Figure 3 (a) and (b) corresponds to image of a laser spot used to excite the sample and the corresponding PL image measured at 20 K. Scale bar, 1 μm. (c) A line profile of PL the excitation laser spot and the PL image shown in Fig. 3(a). A fitted curve based on solving the two-dimensional diffusion equation is also shown.

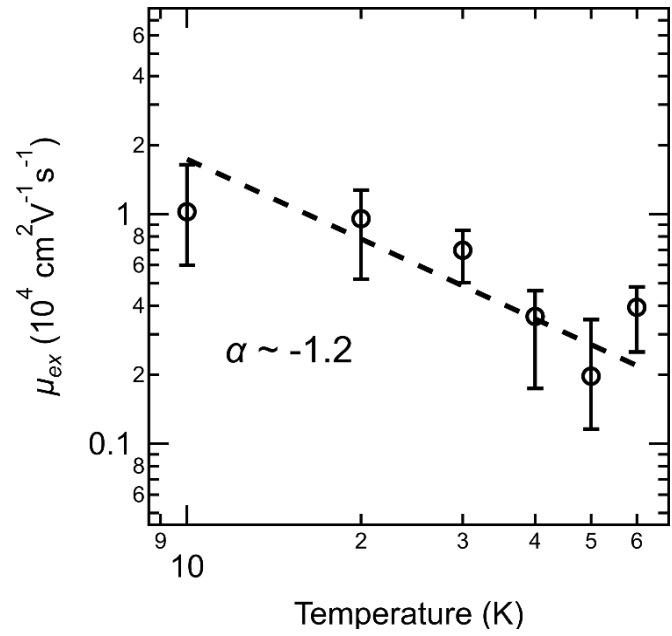

Figure 4 Temperature dependence of exciton mobility, $\mu_{ex}$. The dotted line is a fitted line with equation of $\ln(\mu_{ex})=\ln(T^\alpha)$, where $\alpha$ is -1.2 as shown in the figure.

## S1 Fabrication of hBN-encapsulated MoSe$_2$ flake

We fabricated hBN-encapsulated MoSe$_2$ with a method based on the dry transfer technique. First of all, hBN and MoSe$_2$ (Fig S1a) flakes were respectively exfoliated on a clean SiO$_2$/Si substrate. One of the hBN flakes was picked up with a PMMA/PDMS stamp and then a MoSe$_2$ flake was picked up with the hBN on the PMMA/PDMS stamp. Finally, the stacking structure, hBN/MoSe$_2$, was transferred onto another flake of hBN together with PMMA. The PMMA film on the hBN/MoSe$_2$/hBN was removed with a home-made H$_2$ plasma cleaner. The exposure to H$_2$ plasma does not increase the PL peaks at ~ 1.5 eV, which is usually observed in PL spectra of defective MoSe$_2$ measured at low temperature[1]. We used EB lithography and reactive ion etching to form the sample (CF$_4$ = 30 sccm, O$_2$ = 4 sccm, power: 60 W, pressure: 2.0 Pa, time: 30 s). Au/Cr was deposited to edge of the hBN/MoSe$_2$/hBN sample in order to contact electrically. Unfortunately, this FET device didn't work well at low temperature probably due to the large Schottky barrier.

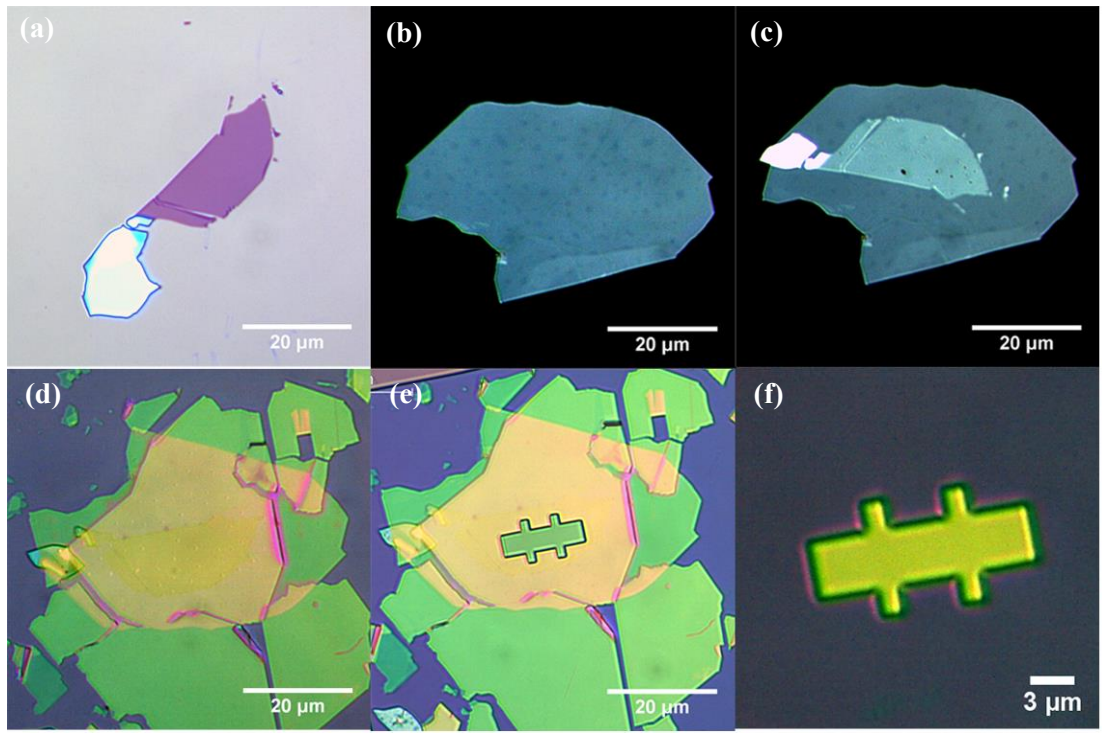

Figure S1(a)-(f) Optical microscope images of a hBN/MoSe$_2$/hBN sample.

## S2 PL measurements and fitting PL peaks by Voigt function

We used excitation wavelength of 550 nm for measurements of PL spectra. Pulsed white light from a super-continuum laser (SuperK EXTREME, NKT Photonics, 40 MHz) is monochromatized by a spectrometer (Princeton Instruments, SP2150) to have the excitation wavelength of 550 nm. The laser beam is focused on the sample surface with 50x objective lens with correction ring (Nikon, CFI L Plan EPI CR, *NA* = 0.7). The sample is put in a continuous flow cryostat (KONTI-Cryostat-Micro, CryoVac) under vacuum condition, and sample temperature is controlled with a temperature measurement control unit (TIC 304-MA, CryoVac). The reflected excitation laser light is blocked with a 600 nm long-pass filter, and PL spectra is measured with a spectrometer (IsoPlane 320, Princeton Instruments) and a charge coupled device (PIXIS 1024B_eXcelon, Princeton Instruments).

Fig. S2 shows a typical PL spectrum measured at 10 K. Two peaks at 1.64 and 1.61 eV are assigned to emission arising from radiative recombination of excitons and trions, respectively. A dashed line in Fig S2 corresponds to a fitting curve with Voigt function, whose parameters are shown in Table S1.

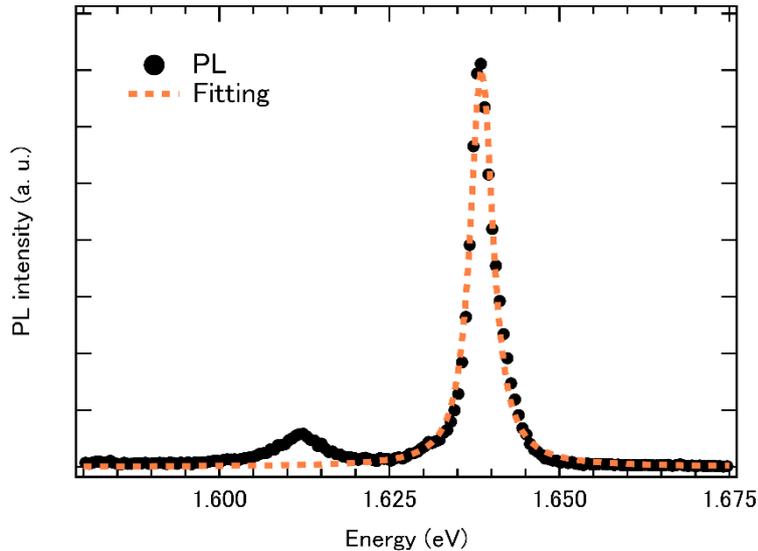

Figure S2 A PL spectrum of a hBN/MoSe$_2$/hBN measured at 10 K. The red dotted line corresponds to a fitted curve with Voigt function.

Table S1 Fitting parameters of Voigt function

|  | Obtained values |
| --- | --- |
| Peak position / eV | $1.638 \pm 1.7 \times 10^{-5}$ |
| Gaussian FWHM / meV | $1.2 \pm 0.23$ |
| Lorentzian FWHM / meV | $3.5 \pm 4.9 \times 10^{-2}$ |

## S3 Varshini's plot

Fig. S3 represents temperature dependence in peak positions of the exciton emission. As shown in the Fig. S3, the peak positions are blue-shifted as temperature decreases, which tendency is consistent to the previously reported results. We fitted the temperature dependence data by Varshini's equation shown below.

$$E_g(T) = E_g(0) - \frac{\alpha T^2}{T + \beta}$$

α and β are parameters and $T$ and $E_g(0)$ represent temperature and the bandgap at $T = 0$ K, respectively. As clearly seen, the Varshini's equation with parameters shown in Table S2 reproduces the observed data well.

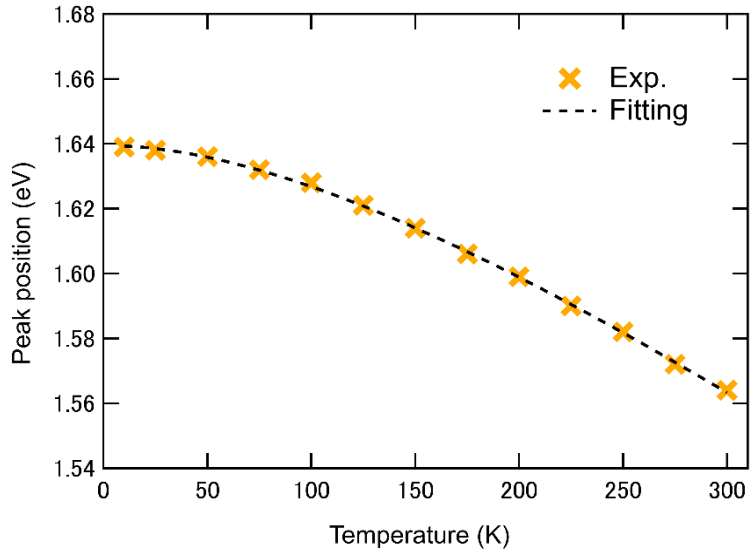

Figure S3 temperature dependence in peak positions of exciton emission. The black dotted line corresponds to a fitted curve with Varshini's equation.

Table S2 Fitting parameters of Varshini's equation

|  | Obtained values |
| --- | --- |
| α / eV K$^{-1}$ | $5.07 \times 10^{-4}$ |
| β / K | $2.99 \times 10^2$ |
| $E_g(0)$ / eV | 1.64 |

## S4 Time-resolved PL

Time-resolved PL data are obtained by time correlated single photon counting (TCSPC) method. Instrument response function (IRF) and PL signal are collected by a photon counter (Becker & Hickl GmbH, ID-100-50-ULN). Fig S5(a)-(f) show the time-resolved PL intensity measured at 10 ~ 60 K. We fitted the measured PL decay with double exponential decay function shown below.

$$I(t) = a_1 \exp\left(-\frac{t}{\tau_1}\right) + (1 - a_1)\exp\left(-\frac{t}{\tau_2}\right)$$

$\tau_1$, $\tau_2$ and $a_1$ are lifetimes and a fitting parameter; obtained parameters are listed in Table S3. Although we used the double exponential decay model, ratio of the long lifetime component in this sample is very small. Fig. S4(g) shows temperature dependence of $\tau_1$, whose error bars were determined based on the parameter errors.

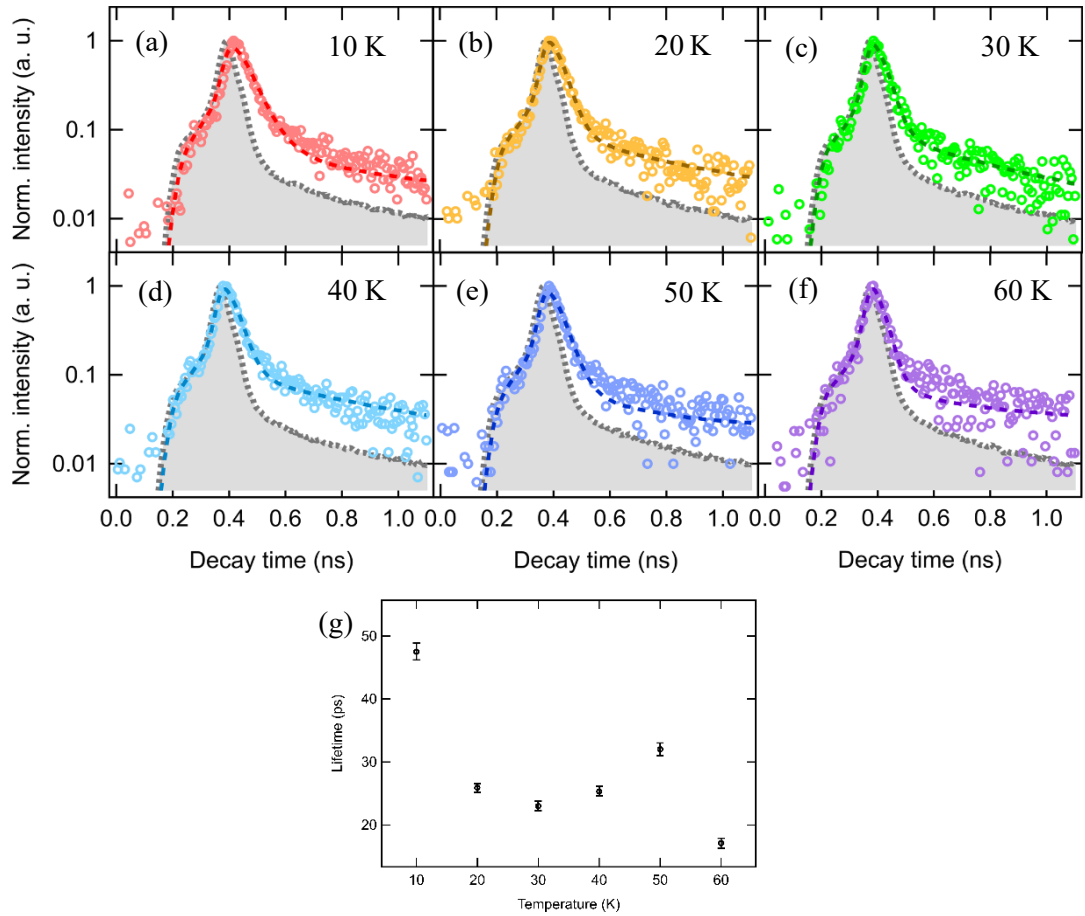

Figure S4(a)-(f) PL decay measured at 10 ~ 60 K. Gray area and plots correspond to IRF and PL signals, respectively. (g) temperature dependence of $\tau_1$

Table S3 Obtained parameters

| T / K | $a_1$ | $\tau_1$ / ps | $\tau_2$ / ns |
|---|---|---|---|
| 10 | 0.994 ± 0.0013 | 47.5 ± 1.3 | 2.44 ± 0.72 |
| 20 | 0.990 ± 0.0013 | 25.9 ± 0.6 | 0.67 ± 0.10 |
| 30 | 0.989 ± 0.0018 | 23.0 ± 0.7 | 0.50 ± 0.09 |
| 40 | 0.987 ± 0.0014 | 25.3 ± 0.7 | 0.68 ± 0.08 |
| 50 | 0.994 ± 0.0007 | 32.0 ± 0.9 | 4.41 ± 0.91 |
| 60 | 0.994 ± 0.0006 | 17.1 ± 0.7 | 3.67 ± 0.57 |

## S5 Solving 2D diffusion equation and fitting of PL profile

Fig. S5(a) represents the residual sums of squares calculated with different values of diffusion constants, $D$. The best $D$ value, which gives the minimum residual sum, was obtained through fitting of the data in Fig. S5(a) with a parabolic function. Fig. S5(b) represents temperature dependence of $D$ values with errors bars, whose ranges correspond to the maximum and the minimum values obtained with different data sets; we have measured PL images, at least, 6 times at each temperature. As clearly seen, $D$ values show weak temperature dependence, which means mobility of exciton increase as temperature decreases.

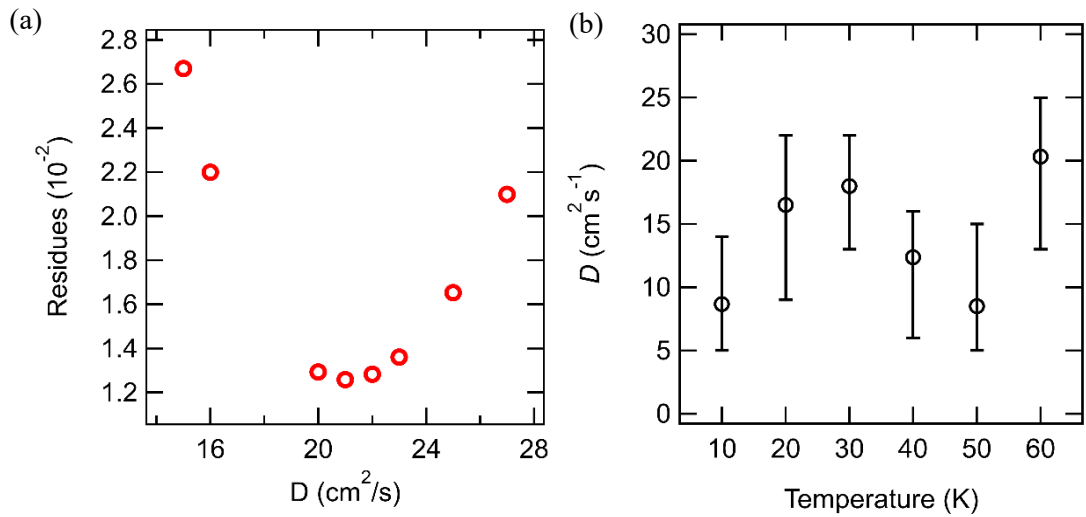

Figure S5 (a) relation between $D$ values and the residual sums of squares. (b) temperature dependence of $D$ obtained.

## S6 $\tau$, D and $\mu_{ex}$ in another sample

Fig. S6(a)-(c) show temperature dependence of lifetime, $D$ values and the corresponding exciton mobility in the other sample. Note that contribution from the long-lived components in the double exponential model is not so small in this sample. As shown in Fig. S6(c), a log-log plot of temperature and exciton mobility also gives a slope of ~1.1.

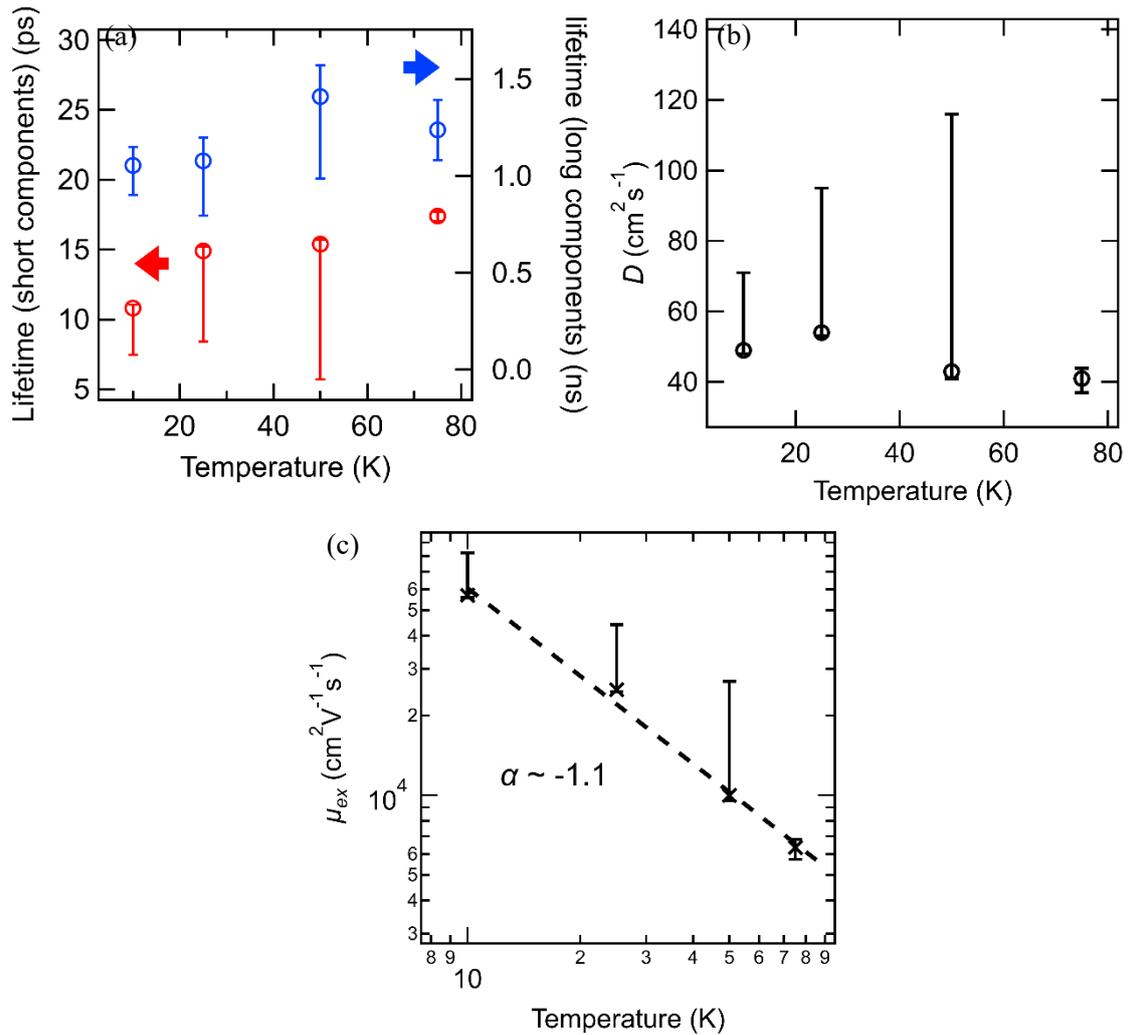

Figure S6a temperature dependence of lifetime, where red and blue plots correspond to short- and long-lived components. (b) and (c) show temperature dependence of diffusion coefficients and exciton mobility.